\documentstyle[12pt]{article}
\begin{document}
\begin{center}
{\bf On Generalized Distributions and Pathways}\\[0.2cm]
{\bf A.M. Mathai}\\[0.1cm]
Department of Mathematics and Statistics, McGill University, 805 Sherbrooke Street West, Montreal, Canada H3A 2K6\\[0.3cm]
{\bf H.J. Haubold}\\[0.1cm]
Office for Outer Space Affairs, United Nations, Vienna International Centre, P.O. Box 500, A-1400 Vienna, Austria\\ 
\end{center}
\noindent
{\bf Abstract.}
The scalar version of the pathway model of Mathai (2005) is shown to be associated with a large number of probability
models used in physics. Different families of densities are listed here, which are all connected through the pathway parameter $\alpha$, generating a distributional pathway. The idea is to switch from one functional form to another through this parameter and it is shown that one can proceed from the generalized type-1 beta family to generalized type-2 beta family to generalized gamma family. It is also shown that the pathway model is available by maximizing a generalized measure of entropy, leading to an entropic pathway, covering the particularly interesting cases of Tsallis statistics (Tsallis, 1988) and superstatistics (Beck and Cohen, 2003). 
\section{Introduction}

When deriving or fitting models for data from physical experiments very often the practice is to take a member from a parametric family of distributions. 
But it is often found that the model requires a distribution with a more specific tail than the ones available from the parametric family, or a situation of right tail cut-off. The model may reveal that  the underlying distribution is in between two parametric families of distributions. In order to create a distributional pathway for proceeding from one functional form to another a pathway parameter $\alpha$ is introduced and a pathway model is created in Mathai (2005). Section 1 defines this pathway model. In Section 2 it is shown that this model enables one to go from a generalized type-1 beta model to a generalized type-2 beta model to a generalized gamma model when the variable is restricted to be positive. More families are available when the variable is allowed to vary over the real line. Corresponding multitudes of families are available when the variable is in the complex domain. This is a pragmatic approach to the distributional pathway. Mathai (2005) deals mainly with rectangular matrix-variate distributions in the real case and the scalar case is the topic of discussion in the current paper because the scalar case is connected to many problems in physics. Further, in Section 3, through a generalized measure of entropy, an entropic pathway can be devised that leads to different families of entropic functionals that produce families of distributions by applying the maximum entropy principle (Mathai and Haubold, 2007). This is a theoretical approach to the distributional pathway. Section 4 draws conclusions from the development of the distributional pathway.
\section{The pathway model in the real scalar case}

For the real scalar case the pathway model for positive random variables is represented by the following probability density, for purely mathematical statistical reasons, and to unify and extend the compilation of statistical distributions in Mathai (1993):
\begin{equation}
f(x)=c\;x^{\gamma -1}[1-a(1-\alpha)x^{\delta}]^ {\frac{\beta}{1-\alpha}},\;\; x>0,
\end{equation}
$a>0,\delta >0, \beta\geq 0,1-a(1-\alpha)x^{\delta}>0,\gamma >0$ where $c$ is the 
normalizing constant and $\alpha$ is the pathway parameter.  The normalizing 
 constant in this case is the following:
\begin{eqnarray}
c&=&\frac{\delta [a(1-\alpha)]^{\frac{\gamma}{\delta}}
\Gamma\left(\frac{\gamma}{\delta}+\frac{\beta}{1-\alpha}+1\right)} 
{\Gamma\left(\frac{\gamma}{\delta}\right)\Gamma\left(\frac{\beta}{1-\alpha}
+1\right)},\;\;\mbox{for}\; \alpha<1 \\
&=& \frac{\delta [a(\alpha-1)]^{\frac{\gamma}{\delta}}\Gamma\left(\frac{\beta}{\alpha-1}\right)}
{\Gamma\left(\frac{\gamma}{\delta}\right)\Gamma\left(\frac{\beta}{\alpha-1}
-\frac{\gamma}{\delta}\right)},\;\;\mbox{for}\;\;\frac{1}{\alpha-1}
-\frac{\gamma}{\delta}>0,\;\;\alpha>1 \\
&=&\frac{\delta\;(a\beta)^{\frac{\gamma}{\delta}}}{\Gamma\left(\frac{\gamma}
{\delta}\right)},\;\mbox{for}\;\;\alpha\rightarrow 1.
\end{eqnarray}
We note here with respect to (1) that in information theory and mathematical statistics, parameter interpretation is less important than in special cases under the interpretation of a physical model or principle (Kaniadakis and Lissia, 2004).\par
\noindent

\subsection{Generalized type-1 beta family emanating\\
from equation (1)}

For $\alpha <1$ or $-\infty<\alpha<1$ the 
model in $(1)$ remains as a generalized type-1 beta model in the real case, where the right
tail is cut-off. In a series of publications, see for example Mathai and Haubold (1988) and 
Haubold and Mathai (2000), we considered various 
modifications 
to the Maxwell-Boltzmann approach to the nuclear reaction rate theory and considered
the situations  of nonresonant reactions and cases of depleted
tail and tail cut-off. It may be observed that all those situations 
are covered by the pathway model in $(1)$. Before proceeding with the properties 
let us look at some special cases first.  For
$\alpha=0$, $a=1$, $\delta=1$ with $\beta$ replaced by $\beta -1$ we have the regular type-1 beta density, namely,
\begin{equation}
f_1(x)=c_1\;x^{\gamma-1}(1-x)^{\beta-1}, 0<x<1,
\end{equation}
where 
$c_1=\frac{\Gamma(\beta+\gamma)}{\Gamma(\gamma)\Gamma(\beta)}$. For  
$a=1,\gamma=1,\delta=1, \beta=1$ we have Tsallis statistics (Tsallis, 1988; Gell-Mann and Tsallis, 2004)
for $\alpha<1$, namely,
\begin{equation}
f_2(x)=c_2\;[1-(1-\alpha)x]^{\frac{1}{1-\alpha}},\;\; 0<x<(1-\alpha)^{-1},
\end{equation}
where $c_1$ is the corresponding normalizing constant. For $\alpha=0,
a=1$ in $(1)$ 
we have the power function densities for $\gamma=1$ and $\beta=0$ respectively, namely,
\begin{equation}
f_3(x)=c_3\;x^{\gamma-1},\;\;0<x<1
\end{equation}
and
\begin{equation}
f_4(x)=c_4\;(1-x^{\delta})^{\beta},\;0<x<1,
\end{equation}
where $c_3,c_4$ are the corresponding normalizing constants. Note that 
Pareto densities (Brouers and Sotolongo-Costa, 2004; Shalizi, 2007) also come from the power function models. Further, if $\gamma=1$, 
$\delta=1$, $a=1,\alpha=0$, $\beta=0$ in $(1)$ 
 we have the uniform density,
\begin{equation}
f_5(x)=1,\;0<x<1.
\end{equation}
Before concluding this section a remark on Bose-Einstein statistics (Ijiri and Simon, 1975; Aragao-Rego et al. 2003) in physics
is in order. Naturally one would expect the Bose-Einstein density to belong to 
the type-1 beta family. If we take the parameters $\gamma=1$ and $\beta=0$ in 
the density in equation $(5)$ then naturally the function goes to
$$g(x)=c^{*}\frac{1}{1-x}$$
and obviously $g(x)$ cannot make it a density in $0\leq x\leq 1$ unless $x$ is 
bounded away from $1$. This can be achieved by a transformation of the 
  type $x=\exp{(-t-wy)}$ for $w>0, e^t>1$. In this case $g(x)$ goes to 
\begin{equation}
g_1(y)=c_1^{*}\frac{w}{e^{t+wy}-1},\; 0\leq y<\infty,\; w>0,\; e^t>1.
\end{equation}
This is nothing but the Bose-Einstein density which is given by
\begin{equation}
f_6(y)=c_6\frac{1}{-1+\exp{(t+wy)}},\; w>0,0\leq y<\infty,\; e^t>1.
\end{equation}\par
\noindent
{\bf Remark 2.1} Thus Bose-Einstein density is a limiting form of a 
type-1 beta density of equation $(5)$ where the variable $x$ is transformed by the above inverse 
transformation where the normalizing constant $c_6$ cannot be obtained through 
gamma functions but it can be evaluated by appealing to partial fractions 
technique and 
then taking logarithms and it can be easily seen to be the following: 
$$c_6=w\left[\ln \left(\frac{e^t}{e^t-1}\right)\right]^{-1}.$$
\noindent
\subsection{Generalized type-2 beta family emanating\\ 
from equation (1)}
 When $\alpha >1$ in $(1)$ we may write $1-\alpha=-(\alpha -1), \alpha >1$ 
so that $f(x)$ assumes the form,
\begin{equation}
f(x)=c\;x^{\gamma-1}[1+a(\alpha-1)x^\delta]^{-\frac{\beta}{\alpha-1}},\;\;x>0
\end{equation}
which is a generalized type-2 beta model for real $x$. Beck and Cohen's  superstatistics 
(Beck and Cohen, 2003; Beck 2006) belongs to this case (12). 
For $\alpha=2,a=1,\delta=1$, $\beta-\gamma>0$ we have the regular type-2 beta density, namely,
\begin{equation}
f_7(x)=c_7\;x^{\gamma-1}(1+x)^{-\beta},\; x>0.
\end{equation}
In $f_7(x)$ for $x=\frac{m}{n}F,\;m,n=1,2,\ldots,\gamma=\frac{m}{2},
\beta=\frac{m}{2}+\frac{n}{2}$ we have the $F$ density or the variance-ratio density, namely,
\begin{equation}
f_8(F)=c_8\;F^{\frac{m}{2}-1}(1+\frac{m}{n}F)^{-(\frac{m}{2}+\frac{n}{2})},\;\;F>0.
\end{equation}
For $\gamma=1,a=1,\delta=1$, $\beta=1$ in (12) we have Tsallis 
statistics (Tsallis, 1988; Gell-Mann and Tsallis, 2004) for $\alpha>1$, namely 
\begin{equation}
f_9(x)=c_9\;[1+(\alpha-1)x]^{-\frac{1}{\alpha-1}},\; x>0.
\end{equation}
For $\delta=2, -\infty<x<\infty$, $a=\frac{1}\nu,\alpha=2, \beta=\frac{\nu+1}{2}$ 
in (12) we have the Student-t density (Gheorghiu and Coppens, 2004), namely
\begin{equation}
f_{10}(x)=c_{10}(1+\frac{x^2}{\nu})^{-\frac{(\nu+1)}{2}},\;\; -\infty<x<\infty.
\end{equation}
In $f_{10}$ for $\nu=1$ we have the Cauchy model, namely,
\begin{equation}
f_{11}(x)=c_{11}(1+x^2)^{-1},\;\;-\infty<x<\infty.
\end{equation}
In $f_7$ if $\gamma=1$, $\beta=1$, $x=e^{\epsilon+\eta y}$ for 
$\eta>0$, $\epsilon\neq 0$, $0\leq y<\infty$ then we have the Fermi-Dirac density (Aragao-Rego et al., 2003)
\begin{equation}
f_{12}(y)=c_{12}[1+\exp{(\epsilon+\eta y)}]^{-1}, 0\leq y<\infty.
\end{equation}\par
\noindent
{\bf Remark 2.2} Observe that $f_{12}$ is  a limiting form of the ordinary type-2 beta model with 
$\beta-\gamma=0$ in $f_7$. In this case the normalizing constant cannot be evaluated
 with the help of gamma functions but $c_7$ can be evaluated through partial 
fractions and then appealing to logarithms.\\ 
In $f_7$ if we transform $x$ to $y$ 
such that $x=e^y$, $-\infty < y <\infty$ then we obtain the 
generalized logistic and related models
\begin{equation}
f_{13}(y)=c_{13}[e^y]^{\gamma}[1+e^y]^{-\beta}, -\infty<y<\infty
\end{equation}
which are applicable in many areas of 
statistical analysis, see for example Mathai and Provost (2006). 
Before concluding this section one more remark is in order.\par
\noindent
{\bf Remark 2.3} For the generalized type-2 beta model, that is the 
model in (1) or the model in (12) for $\alpha>1$, $x$ and $\frac{1}{x}$ belong to the same family
of distributions. Hence we could have replaced $x^{\delta}$ by 
$x^{-\delta}$ in the densities in equations (13) to (19) then we would obtain 
the corresponding additional families of densities.
\subsection{Generalized gamma family emanating\\ 
from equation (1)}

When $\alpha\rightarrow 1$ the forms in $(1)$ for $\alpha <1$ and for $\alpha>1$ reduces 
to
\begin{equation}
f(x)=c\;x^{\gamma-1} e^{-a\beta\;x^\delta},\;\; x>0.
\end{equation}
This includes generalized gamma, gamma, exponential, chisquare, Weibull, 
Maxwell-Boltzmann, Rayleigh, and related models (Mathai, 1993).  The model
in equation (20) is generally known as the generalized gamma model. For 
$\gamma=1, \delta=2, -\infty<x<\infty$, $a=1$, in (20) we have the Gaussian 
density or error curve,
\begin{equation}
f_{14}(x)=c_{14}\exp{(-\beta x^2)}, -\infty<x<\infty.
\end{equation}
For $\gamma=\delta, a=1$ in (20) we have the Weibull density, given by
\begin{equation}
f_{15}(x)=c_{15}x^{\delta-1}e^{-\beta x^\delta},\;\; x>0.
\end{equation}
For $\delta=1$, $a=1$ in (20) we have the gamma density, given by
\begin{equation}
f_{16}(x)=c_{16}x^{\gamma-1}e^{-\beta x},\;\; x>0.
\end{equation}
For $\gamma=\frac{n}{2}$ and $\beta=\frac{1}{2}$ in $f_{16}$ we have the chisquare 
density with $n$ degrees of freedom, given by
\begin{equation}
f_{17}(x)=c_{17}x^{\frac{n}{2}-1}e^{-x/2},\; x>0.
\end{equation}
For $\gamma=n,\beta=\frac{n}{2\sigma^2}, \sigma>0, n=1,2,..., \delta=2, a=1$ in 
(20) we have the chi density, given by
\begin{equation}
f_{18}(x)=c_{18}x^{n-1}e^{-\frac{nx^2}{2\sigma^2}},\; x>0.
\end{equation} 
For $\gamma=2,\ldots$ in $f_{16}$ we have the Erlang density, given by
\begin{equation}
f_{19}=c_{19}x^{p-1}e^{-\beta x}, p=2,\ldots,\; x>0.
\end{equation}
For $p=1$ in $f_{19}$ we have the exponential density, given by
\begin{equation}
f_{20}(x)=c_{20}e^{-\beta x},\; x>0.
\end{equation}
For $\delta=2,\gamma=3,a=1$ in (20) we have the Maxwell-Boltzmann density, 
given by
\begin{equation}
f_{21}(x)=c_{21}x^2 e^{-\beta x^2},\; x>0.
\end{equation}
For $\delta=2,\gamma=2,\beta=\frac{1}{2\sigma^2}, \sigma>0, a=1$ in (20) 
we have the Rayleigh density (Tirnakli et al., 1998), given by
\begin{equation}
f_{22}(x)=c_{22}x e^{-\frac{x^2}{2\sigma^2}}, x>0.
\end{equation}\par
\noindent
{\bf Remark 2.4} Observe that in (20) if we replace $\delta>0$ 
by $-\delta, \delta>0$ still the family belongs to the generalized gamma family. 
Hence a sequence of densities are available by replacing $x$ by $\frac{1}{x}$
in (20).\par
\noindent
{\bf Remark 2.5} If $x$ is replaced by $|x|$ for $-\infty<x<\infty$ in
the models (1), (12) and (20) we obtain a series of other densities which 
will also include double exponential or the Laplace density, among others.\par
\noindent
{\bf Remark 2.6} In many practical problems there may be a location 
parameter or the variable may be located at a point different from zero. All 
such cases can be covered by replacing $x$ in the pathway model in $(1)$ by
$(x-b)$ when $x-b>0$ or by $|(x-b)|$ when $-\infty<x<\infty$, where $b$ 
is a constant, called the location parameter.\\  

Observe that in (12) and (20), $\frac{1}{x}$ also belongs to the same 
family of densities and hence in (12) and (20) one could have also taken 
$x^{-\delta}$ with $\delta>0$.
\bigskip
\noindent
\section{Pathway model from a generalized entropy measure}
A generalized entropy measure of order $\alpha$, explored by Mathai and Haubold (2007),
is a generalization of Shannon entropy and it is  a variant of the 
generalized entropy of order 
$\alpha$ in Mathai and Rathie (1975). In the discrete case the measure is the 
 following: 
Consider a multinomial population $P=(p_1,\ldots, p_k)$,$\;p_i\geq 0$,$\; i=1,\ldots, k$, 
$p_1+\ldots +p_k=1$. Define the function
\begin{equation}
M_{k,\alpha}(P)=\frac{\sum_{i=1}^k p_i^{2-\alpha}-1}
{\alpha-1},\;\alpha\neq 1,\;\;-\infty<\alpha<2 
\end{equation}
\begin{equation}
\lim_{\alpha\rightarrow 1}M_{k,\alpha}(P) =-\sum_{i=1}^k p_i\ln p_i
=S_k(P)
\end{equation}
by using L'Hospital's rule. In this notation $0\ln 0$ is taken as zero when 
any $p_i=0$. Thus (30) is a generalization of Shannon entropy $S_k(P)$ as seen 
from (31). Note that (30) is a variant of Harvda-Charv\'at entropy 
$H_{k,\alpha}(P)$ and Tsallis entropy $T_{k,\alpha}(P)$ where
\begin{equation}
H_{k,\alpha}(P)=\frac{\sum_{i=1}^kp_i^\alpha-1}{2^{1-\alpha}-1},\;\alpha\neq 1,\;\alpha>0
\end{equation}
and
\begin{equation}
T_{k,\alpha}(P)=\frac{\sum_{1=1}^kp_i^\alpha-1}{1-\alpha},\;\;\alpha\neq 1,\;\;\alpha>0.
\end{equation}
We will introduce another measure associated with (31) and parallel to 
R\'enyi entropy $R_{k,\alpha}$ in the following form:
\begin{equation}
M_{k,\alpha}^*(P)=\frac{\ln\left(\sum_{i=1}^k p_i^{2-\alpha}\right)}{\alpha-1},\;\;\alpha\neq 1,-\infty<\alpha<2.
\end{equation}
R\'enyi entropy is given by
\begin{equation}
R_{k,\alpha}(P)=\frac{\ln\left(\sum_{i=1}^k p_i^\alpha \right)}{1-\alpha},\;\; \alpha\neq 1,\;\alpha>0.
\end{equation}
It will be seen later that the form in (30) is amenable to the pathway model.\par 
\medskip
\noindent
\subsection{Continuous analogue}

The continuous analogue to the measure in (30) is the following:
\begin{eqnarray}
M_\alpha(f)&=& \frac{\int_{-\infty}^\infty[f(x)]^{2-\alpha}dx-1}
{\alpha-1}\\
&=&\frac{\int_{-\infty}^{\infty}[f(x)]^{1-\alpha}f(x)dx-1}
{\alpha-1}=\frac{E[f(x)]^{1-\alpha}-1}{\alpha-1},\;\;\alpha\neq 1,\;\alpha <2,\nonumber
\end{eqnarray}
where $E[\cdot]$ denotes the expected value of $[\cdot]$. 
Note that when $\alpha =1,\\
E[f(x)]^{1-\alpha}=E[f(x)]^0=1.$\\

It is easy to see that the generalized entropy measure in (30) is connected to Kerridge's
 ``inaccuracy" measure (Kerridge, 1961). The generalized inaccuracy measure is 
$E[q(x)]^{1-\alpha}$ where the experimenter has assigned $q(x)$ for the true
 density $f(x)$, where $q(x)$ could be an estimate of $f(x)$ or $q(x)$ could
 be coming from observations. Through disturbance or distortion if the 
experimenter assigns $[f(x)]^{1-\alpha}$ for $[q(x)]^{1-\alpha}$ then
 the inaccuracy measure is $M_{\alpha}(f)$ of (36). 
\subsection{Distributions with maximum generalized entropy}

Among all densities, which one will give a maximum value for $M_{\alpha}(f)$ in equation (36)? 
Consider all possible functions $f(x)$ such that $f(x)\geq 0$ for all $x$, 
$f(x)=0$ outside $(a,b)$, $a<b$, $f(a)$ is the same for all such $f(x)$, 
$f(b)$ is the same for all such $f$, 
 $\int_{a}^{b}f(x)dx<\infty$.  Let $f(x)$ be a continuous function of $x$ with continuous 
derivatives in $(a,b)$. Let us maximize $\int_{a}^{b}[f(x)]^{2-\alpha}dx$ 
for fixed $\alpha$ and over all functional $f$, under the conditions that the 
following two moment-like expressions be fixed quantities:
\begin{equation}
\int_a^b x^{(\gamma-1)(1-\alpha)}f(x)dx=\mbox{given, and}
\int_a^b x^{(\gamma-1)(1-\alpha)+\delta}f(x) dx=\mbox{given}
\end{equation}
for fixed $\gamma>0$ and $\delta>0$. Consider
\begin{equation}
U=[f(x)]^{2-\alpha}-\lambda_1 x^{(\gamma-1)(1-\alpha)}f(x)+\lambda_2 x^{(\gamma-1)(1-\alpha)+\delta}f(x),\;\alpha<2,\;\alpha\neq 1
\end{equation}
where $\lambda_1$ and $\lambda_2$ are Lagrangian multipliers. Then the Euler equation is the following:
\begin{eqnarray}
\frac{\partial U}{\partial f}= 0 &\Rightarrow& (2-\alpha)[f(x)]^{1-\alpha}-\lambda_1 x^{(\gamma-1)(1-\alpha)}+\lambda_2x^{(\gamma-1)(1-\alpha)+\delta}=0\nonumber\\
&\Rightarrow& [f(x)]^{1-\alpha}=\frac{\lambda_1}{(2-\alpha)} x^{(\gamma-1)(1-\alpha)}[1-\frac{\lambda_2}{\lambda_1}x^{\delta}]\\
&\Rightarrow& f(x)=c\;x^{\gamma-1}[1-\eta(1-\alpha)x^{\delta}]^{\frac {1}{1-\alpha}}
\end{eqnarray}
where $\lambda_1/\lambda_2$ is written as $\eta(1-\alpha)$ with $\eta>0$ 
such that $1-\eta (1-\alpha)x^{\delta}>0$ since $f(x)$ is assumed to be 
non-negative. By using the conditions (37) and (39) we can determine $c$ and $\eta$.
 When the range of $x$ for which $f(x)$ is nonzero is $(0,\infty)$ and when 
$c$ is a normalizing constant, then (40) is the pathway model of Mathai 
(2005) in the scalar case where $\alpha$ is the pathway parameter. When 
$\gamma=1,\delta=1$ in (40) then (40) produces the power law. The form in (39) for 
various values of $\lambda_1$ and $\lambda_2$ can produce all the four forms
$$\alpha_1x^{\gamma-1}[1-\beta_1(1-\alpha)x^{\delta}]^{-\frac{1}{1-\alpha}},\;
\alpha_2x^{\gamma-1}[1-\beta_2(1-\alpha)x^{\delta}]^{\frac{1}{1-\alpha}}\mbox{for}\alpha<1$$
and
$$\alpha_3 x^{\gamma-1}[1+\beta_3(\alpha-1)x^{\delta}]^{-\frac{1}{\alpha-1}},\;\alpha_4x^{\gamma-1}[1+\beta_4(\alpha-1)x^{\delta}]^{\frac{1}{\alpha-1}} \mbox{for}\alpha>1$$
with $\alpha_i,\beta_i>0,i=1,2,3,4$. But out of these, the second and the third
 forms can produce densities in $(0,\infty)$. The first and fourth will not be 
converging. When $f(x)$ is a density in (40), what is the normalizing constant 
$c$? We need to consider three cases of $\alpha<1,\alpha>1$ and 
$\alpha\rightarrow 1$. This $c$ is already evaluated in section 1.\par
\noindent
{\bf Remark 3.1} In Mathai and Haubold (2007) further results are provided for the pathway
model associated with Tsallis' entropy, fractional calculus, Mittag-Leffler functions, and distributions with one of the parameters having a prior distribution of its own, giving rise to superstatistics (Beck and Cohen, 2003; Beck, 2006).
\bigskip
\noindent
\section{Conclusions}
Based on the pathway model developed by Mathai (2005) we derived a {\it distributional pathway} proceeding from the generalized type-1 beta family to generalized type-2 beta family in eq. (12) to generalized gamma family in eq. (20). This distributional pathway encompasses, among others, the distributions of Maxwell-Boltzmann and Tsallis (q-exponential function, see Gell-Mann and Tsallis 2004; Tsallis, 2004) that are fundamental to Boltzmann-Gibbs statistical mechanics and its nonextensive generalization by Tsallis, respectively, as well as Bose-Einstein and Fermi-Dirac distributions. Subsequent to the distributional pathway, an {\it entropic pathway}, emanates from eq. (1) by maximum principle applied to the generalized entropic form of order $\alpha$ in (30), covering entropic functionals of Shannon, Boltzmann-Gibbs, R\'enyi, Tsallis, and Harvda-Charv\'at (Mathai and Haubold, 2007). The results in this paper are a contribution to the on-going debate in the physical literature concerning the generalization of Boltzmann-Gibbs statistical mechanics through a one-parameter generalization of Boltzmann-Gibbs entropy, known as Tsallis entropy contained in the seminal paper of Tsallis (1988), whose form is equivalent to that of Harvda-Charv\'at, and which opened the door to generalizations of Boltzmann-Gibbs statistical mechanics.
\bigskip
\noindent
\begin{center}
{\bf References}
\end{center}

\noindent
Aragao, H.H., Soares, D.J., Lucena, L.S., da Silva, L.R., Lenzi, E.K., and Fa, K.S. (2003). Bose-Einstein and Fermi-Dirac distributions in nonextensive Tsallis statistics: an exact study. {\it Physica}, {\bf A317}, 199-208.\par
\smallskip
\noindent
Beck, C. (2006). Stretched exponentials from superstatistics. 
{\it Physica}, {\bf A 365}, 96-101.\par
\smallskip
\noindent
Beck, C. and Cohen, E.G.D. (2003). Superstatistics. {\it Physica}, {\bf A322},
267-275.\par
\smallskip
\noindent
Brouers, F. and Sotolongo-Costa, O. (2004). Prior measure for nonextensive entropy. {\it arXiv, cond-mat/0410738 v1}.\par
\smallskip
\noindent
Gell-Mann, M. and Tsallis, C. (Eds.)(2004). {\it Nonextensive Entropy: Interdisciplinary Applications}. Oxford University Press, New York.\par
\smallskip
\noindent
Gheorghiu, S. and Coppens, M.-O. (2004). Heterogeneity explains features of "anomalous" thermodynamics and statistics. {\it Proceedings of the National Academy of Sciences USA}, {\bf 101}, 15852-15856.\par
\smallskip
\noindent
Haubold, H.J. and Mathai, A.M. (2000). The fractional kinetic equation and 
thermonuclear functions. {\it Astrophysics and Space Science}, {\bf 273}, 
53-63.\par
\smallskip
\noindent
Ijiri, Y. and Simon, H.A. (1975). Some distributions associated with Bose-Einstein statistics. {\it Proceedings of the National Academy of Sciences USA}, {\bf 72}, 1654-1657.\par
\smallskip
\noindent
Kaniadakis, G. and Lissia, M. (2004). Editorial for the proceedings of the NEXT2003 conference on news and expectations in thermostatistics, {\it arXiv, cond-mat/0409615v1}.\par
\smallskip
\noindent
Kerridge, D.F. (1961). Inaccuracy and inference. {\it Journal of the Royal 
Statistical Society, Series B}, {\bf 23}, 184-194.\par
\smallskip
\noindent
Mathai, A.M. (1993). {\it A Handbook of Generalized Special Functions for Statistical and Physical Sciences}, Clarendon Press, Oxford.\par
\smallskip
\noindent
Mathai, A.M. (2005). A pathway to matrix-variate gamma and normal densities. {\it 
Linear Algebra and Its Applications}, {\bf 396}, 317-328.\par
\smallskip
\noindent
Mathai, A.M. and Haubold, H.J. (1988). {\it Modern Problems in Nuclear 
and Neutrino Astrophysics}, Akademie-Verlag, Berlin.\par
\smallskip
\noindent
Mathai, A.M. and Haubold, H.J. (2007). Pathway model, Tsallis statistics, 
superstatistics, and a generalized measure of entropy. {\it Physica A}, {\bf 375}, 110-122.\par
\smallskip
\noindent
Mathai, A.M. and Provost, S.B. (2006). On q-logistic and related distributions, 
{\it IEEE Transactions on Reliability}, {\bf 55(2)}, 237-244.\par
\smallskip
\noindent
Mathai, A.M. and Rathie, P.N. (1975). {\it Basic Concepts in Information Theory 
and Statistics: Axiomatic Foundations and Applications}, Wiley Halsted, New York
 and Wiley Eastern, New Delhi.\par
\smallskip
\noindent
Shalizi, C.R. (2007). Maximum likelihood estimation for q-exponential (Tsallis) distributions. {\it arXiv, math.ST/0701854v2}\par
\smallskip
\noindent
Tirnakli, U., Buyukkilic, F., and Demirhan, D. (1998). Some bounds upon the nonextensivity parameter using the approximate generalized distribution functions. {\it Physics Letters}, {\bf A245}, 62-66.\par
\smallskip
\noindent
Tsallis, C. (1988). Possible generalization of Boltzmann-Gibbs statistics. {\it Journal of Statistical Physics}, {\bf 52}, 479-487.\par
\smallskip
\noindent
Tsallis, C. (2004). What should a statistical mechanics satisfy to reflect 
nature? {\it Physica}, {\bf D193}, 3-34.
\end{document}